\documentclass[11pt]{article}

\usepackage{amsmath, amssymb, amsfonts, amsthm, array, authblk, 
            color, enumerate, enumitem, graphicx, epstopdf, epsfig, epsf, 
            float, geometry, helvet, ifthen, latexsym, lscape, multicol, 
            natbib, overpic, setspace, subfigure, titlesec, ulem, wrapfig, mathtools,bbm, titlecaps, xcolor, soul} 
\usepackage[colorlinks,citecolor=blue,urlcolor=blue]{hyperref}
\usepackage{macros}
\usepackage{algorithm2e}
\RestyleAlgo{ruled}

\title{Adaptive Functional Principal Component Analysis}
\author[1,*]{\'Angel Garc\'ia de la Garza}
\author[2]{Britton Sauerbrei}
\author[3]{Adam Hantman}
\author[4]{Jeff Goldsmith}

\affil[1]{Division of Biostatistics, Albert Einstein College of Medicine}
\affil[2]{Department of Neurosciences, Case Western Reserve University}
\affil[3]{Neuroscience Center, University of North Carolina}
\affil[4]{Department of Biostatistics, Columbia University}
\affil[*]{\it angel.garciadelagarza@einsteinmed.edu}

\begin{document}

\maketitle

\setstretch{1.8}

\begin{abstract}

We introduce Adaptive Functional Principal Component Analysis, a novel method to capture directions of variation in functional data that exhibit sharp changes in smoothness.  We first propose a new adaptive scatterplot smoothing technique that is fast and scalable, and then integrate this technique into a probabilistic FPCA framework to adaptively smooth functional principal components. Our simulation results show that our approach is better able to model functional data with sharp changes in smoothness compared to standard approaches. We are motivated by the need to identify coordinated patterns of brain activity across multiple neurons during reaching movements prompted by an auditory cue, which enables understanding of the dynamics in the brain during dexterous movement. Our proposed method captures the underlying biological mechanisms that arise in data obtained from a mouse experiment focused on voluntary reaching movements, offering more interpretable activation patterns that reflect sharp changes in neural activity following the cue. We develop accompanying publicly available software for our proposed methodology, along with implementations to reproduce our results. 

\end{abstract}

Key Words: Adaptive Ridge, Adaptive Smoothing, Functional Data Analysis, Dimension Reduction, Neuron Spike Data


\section{Introduction}

Functional data analysis (FDA) is concerned with settings where observations made on study units are functions measured over time, space, or another continuum. Methods for analyzing such data borrow information across adjacent points in the functions' domain and in that way differ from multivariate approaches \citep{ramsay_functional_2005}. For example, functional principal component analysis (FPCA) is a dimension reduction technique that identifies a set of orthogonal functional principal components (FPCs) that are continuous and smooth. A central consideration in FDA is how to model smoothness most appropriately when conducting an analysis. Whether through a basis expansion, the structure of smoothness-enforcing penalties, or some other mechanism, FDA methods, including FPCA,  typically make the implicit assumption that there is a similar degree of smoothness across the functional domain. When the underlying smoothness in the data fluctuates, this will lead to models that under- and over-smooth over different sections of the data domain. 

\begin{figure}[htb]
  \centering
     \begin{tabular}{cc}
       \includegraphics[width= 1\textwidth]{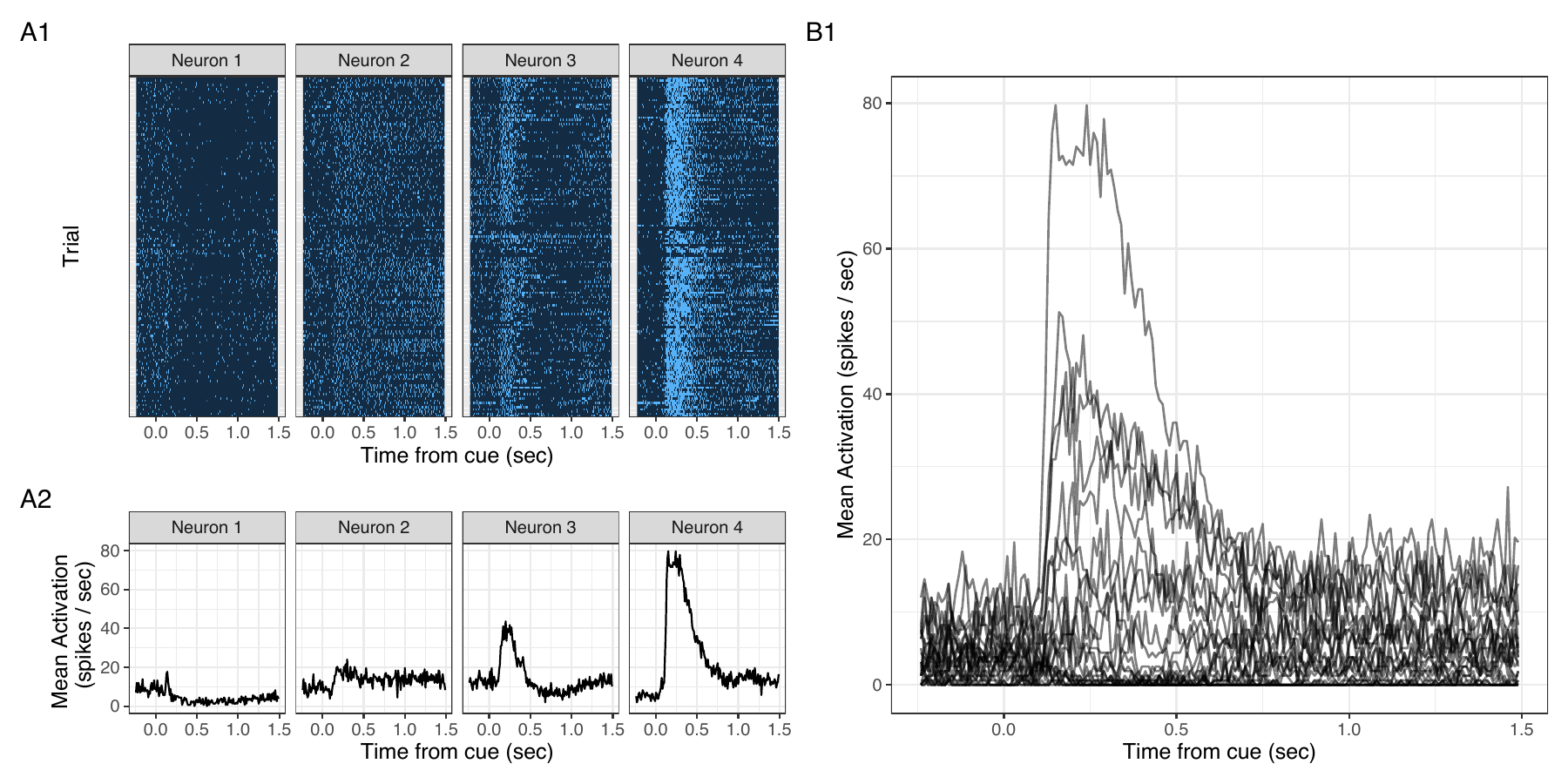} 
     \end{tabular}
     \caption{Panel A1 displays lasagna plots of the activation of four example neurons over a 1.75 seconds interval beginning 0.25 seconds before an auditory cue and 157 trials. Light blue indicates neuron is active. Panel A2 displays the average activation of the same four neurons. Panel B plots average activation for all 25 neurons in our sample.}
    \label{fig:figure1}
\end{figure}

Samples of curves that exhibit locally-varying degrees of smoothness arise regularly. In each trial of the experiment that motivates our work, a trained mouse reaches for a food pellet after hearing an auditory cue while continuous measurements of spike activity in 25 neurons on the motor cortex are recorded using silicon probes \citep{sauerbrei2020cortical}. Before the cue, the mouse's motor cortex is at rest; the auditory cue triggers an immediate response in the motor cortex and, subsequently, a voluntary reach. In the later stages of the reach, neural activation declines slowly and smoothly. Figure \ref{fig:figure1} summarizes our data. Panel A1 shows binary neural activation across the reaching movement recorded in 10ms windows for each of 157 trials in four representative neurons. Panel A2 shows the trial-averaged neuron-specific means measured in spikes per second for the same four neurons, which reflect these neurons' typical activation during the reaching experiment, and Panel B1 shows the trial-averaged activation for all 25 neurons. The neural processes exhibit sharp changes in activity immediately following the auditory cue but are comparatively smooth during the remainder of the observation window. 

Our scientific goal is to identify the activation patterns that emerge across neurons during voluntary motor behavior. These activation patterns summarize the neural behavior of the motor cortex, and are thus more informative than the firing rate in single neurons for understanding cortical activation and involvement in the generation of voluntary movement. Patterns derived using state-of-the-art methods for dimension reduction fail to capture the non-constant smoothness in these data and thus do not reflect the underlying biological behavior. We therefore propose an innovative approach to dimension reduction for functional data in which the level of smoothness varies locally. We first develop a new technique for locally adaptive scatterplot smoothing and then incorporate that into the estimation of FPCs using a penalized likelihood framework. Importantly, our approach estimates all necessary tuning parameters without the need for a computationally expensive cross-validation procedure. Simulations indicate that our proposed adaptive FPCA method outperforms competing approaches when the data generating mechanism includes non-constant degrees of smoothness. Although it is not necessary for our motivating data, our approach and software implementation allows sparse and irregular grids for observed functional data. Applying our method to the motivating data leads to interpretable activation patterns across the motor cortex, clearer scientific conclusions, and robust fits to observed data.

The rest of this manuscript is organized as follows. Section 2 provides a review of the relevant literature. Section 3 contains subsections reviewing a penalized likelihood approach to FPCA, introducing our method for adaptive scatterplot smoothing, and developing the adaptive FPCA model specification. Section 4 presents simulations designed to compare our approach to existing techniques, and Section 5 contains the application of our method's to our motivating neuron spike data. We close with a discussion in Section 6.


\section{Literature Review}
\label{sec:litrev}

Our contributions build on prior work in FPCA, adaptive scatterplot smoothing, and adaptive ridge penalties; we review the relevant literatures in Sections 2.1, 2.2, and 2.3, respectively.


\subsection{Functional Principal Component Analysis}
\label{subsec:fpca}

Because we are primarily interested in the role of smoothing in dimension reduction, we focus our review on existing approaches to smoothing in FPCA. FPCs are frequently obtained through an eigendecomposition of the covariance operator of functional observations $Y(t)$, defined as $\Sigma(u,v) = \text{Cov}\! \left[Y(u),Y(v) \right]$ \citep{besse1986principal}. Early approaches to FPCA smoothed observed curves before estimating the covariance operator \citep{ramsay1991some} or implemented a non-functional PCA and smoothed the resulting components to obtain FPCs \citep{rice1991estimating, pezzulli1993some, silverman1996smoothed}. More recently, it has been common to smooth an empirical covariance surface estimated from observed data and then decompose the result. Examples of this general approach include bivariate kernel or kernel-based approaches  \citep{boente2000kernel, yao2005functional, hall2006properties}; penalized tensor product splines \citep{di2009multilevel,goldsmith2013corrected}; and fast bivariate P-splines \citep{xiao2016fast,xiao2018fast}. 

Methods based on probabilistic principal component analysis \citep{tipping_probabilistic_1999} estimate FPCs by maximizing a likelihood rather than estimating, smoothing, and decomposing a covariance operator. As a result, they may be appealing for data observed over sparse or irregular grids or when the dimension of the observation grid makes smoothing an empirical covariance computationally challenging. Probabilistic FPCA methods include the latent-factor approach for Gaussian data by \citet{james2000principal};  the variational Bayesian approach for binary and count data developed by \citet{van2009bayesian}; and the Bayesian generalized multilevel FPCA extension developed by \citet{goldsmith2015generalized}. These approaches estimate FPCs directly and often include explicit penalties to enforce smoothness on the results. In contrast to our proposed methods, however, neither covariance-based nor probabilistic approaches allow for locally-varying degrees of smoothness.


\subsection{Adaptive Scatterplot Smoothing}
\label{subsec:asmooth}

We next discuss techniques for adaptive scatterplot smoothing in non-functional settings. Scatterplot smoothing considers observations $\Big\{(t_j, y_j): j = \{1,\dots,J\} \Big\}$ and focuses on estimating $E[y] = f(t)$ as a smooth function of $t$.  The goal is to estimate $f(\cdot)$ in a way that balances the goodness of fit to the data against the complexity of $\hat{f}(\cdot)$. A common approach is to penalize the outcome likelihood using the integrated squared second derivative $\int f^{\prime \prime}(t)^2 dt$; a parameter $\lambda$ tunes the relative contribution of the likelihood and the penalty terms in the objective function. The literature on scatterplot smoothing is too vast to thoroughly review here, and instead we will focus narrowly on spline-based methods. As a starting point, we assume that $f(\cdot)$ is expressed as a set of spline basis functions and respective coefficients. The ridge penalty \citep{hoerl1970ridge, brown1980adaptive} is a useful tool to implement non-adaptive scatterplot smoothing, as the integrated squared second derivative penalty is can be expressed as an $L_2$ penalty on the spline coefficients \citep{wood2001mgcv}. An appropriate mixed model will yield the same objective function and simplify the estimation of tuning parameters; this relationship underlies many techniques in scatterplot smoothing and functional data analysis (see \citet{ruppert_semiparametric_2003} and \citet{hodges2013richly}). 

The non-adaptive smoothing penalty will locally under- or over-smooth when the degree of curvature of $f(t)$ varies over $t$ \citep{wahba1990spline, gu1990adaptive}. Broadly, approaches in adaptive smoothing either replace the tuning parameter $\lambda$ with a penalty function $\lambda(t)$ defined over the domain of the data and directly estimate it, or individually penalize each coefficient using a smoothing splines framework. \citet{ruppert_spatially-adaptive_2000} proposed finding a set of penalty constants using a GCV criterion. \citet{baladandayuthapani_spatially_2005} proposed a Bayesian hierarchical model in which spline coefficients have priors with unique variances and parameters are estimated using Markov chain Monte Carlo (MCMC). \citet{krivobokova_fast_2008} developed a fast implementation using a similar hierarchical model in which spline coefficients have a smooth variance structure modeled using a truncated polynomial expansion. \citep{pintore2006spatially} proposed fitting $\lambda(t)$ as a piece-wise function fitted with a kernel and lastly \citet{liu2010data} estimates $\lambda(t)$ as a step-function fitted using an AIC-like criterion. In Section~\ref{subsec:adaptivesmooth_scatterplot}, we derive an explicit link between fitting a form of $\lambda(t)$ and fitting a set of penalty constants, drawing from the adaptive ridge literature to introduce an efficient estimation algorithm.


\subsection{Adaptive Ridge Penalty}
\label{subsec:adaptiveridge}

A  standard ridge penalty on coefficients $\Big\{\beta_p: p \in \{1,\dots,P\}\Big\}$, takes the form $\lambda \sum_{p=1}^P \beta_p^2$, where $\lambda$ is a tuning parameter. This regularizes the coefficient estimates and has a closed form solution for fixed values of $\lambda$ and outcomes that have a Gaussian distribution \citep{hoerl1970ridge, brown1980adaptive}. The adaptive ridge (AR) is a modification of the ridge that assigns a different tuning parameter to each of the coefficients through the penalty $ \lambda \sum_{p=1}^P w_p^2 \beta_p^2$ \citep{grandvalet1998least, canu1999outcomes}. This penalty is ``adaptive'' in the same sense as the adaptive LASSO, in that each coefficient has a unique tuning parameter or weight. AR penalties have been implemented using iterative algorithms that alternate between updating the tuning weights $\Big\{w_p: p \in \{1,\dots,P\}\Big\}$ based on the current estimates of the coefficients, updating the coefficients $\Big\{\beta_p: p \in \{1,\dots,P\}\Big\}$ given the current weights, and selecting a tuning parameter $\lambda$ based on cross-validation or an information criterion such as BIC \citep{frommlet2016adaptive, dai2020broken}. Our work on adaptive scatterplot smoothing and adaptive FPCA, meanwhile, casts the adaptive squared second derivative penalty in terms of an AR penalty, with all tuning parameters estimated analogously to previous mixed model approaches and avoiding computationally expensive cross-validation.


\section{Methods} 
\label{sec:methods}

In this section, we propose new methods to identify patterns of variation in functional data that exhibit sharp changes at some locations in the functional domain but are smoothly varying elsewhere, a structure that is exemplified by our motivating data. First, we briefly outline the technical details of an existing likelihood-based method for FPCA that is not adaptive. We next develop a novel approach to adaptive scatterplot smoothing, and then introduce a new FPCA technique that incorporates our adaptive scatterplot smoothing approach into a likelihood-based FPCA framework and is able to capture sharp changes in smoothness in patterns that underlie observed data.

\subsection{Likelihood-based non-adaptive FPCA}
\label{subsec:fpca_model}

Define $X_i(t)$ to be a set of functions measured over $t \in \mathcal{T}$ for observations $1 \leq i \leq I$ with common mean $\mu(t)$ and covariance operator $\Sigma\left(u,v\right) = \text{Cov}\!\left[X(u), X(v)\right]$. Mercer's theorem provides a decomposition of the covariance operator based on eigenvalues and eigenfunctions; a Kosambi-Karhunen-Lo\`{e}ve (KKL) expansion of functions $X_i(t)$ using these is given by

\begin{equation}
\label{eq:fpca_conceptual}
  E\left[X_i(t)\right] = \mu(t) + \sum_{k=1}^{\infty} \xi_{ik}  \phi_k(t)
\end{equation}

\noindent
where $\boldsymbol{\Phi}(t)=\Big\{\phi_{k}(t): k \in \mathbb{Z}^{+}\Big\}$ are orthonormal eigenfunctions, $\boldsymbol{\eta}=\Big\{\eta_{k}: k \in \mathbb{Z}^{+}\Big\}$ are the corresponding eigenvalues, and scores $\xi_{ik}=\int_{0}^{1}\left[X_i(t)-\mu(t)\right] \phi_{k}(t) d t$ are uncorrelated random variables with mean zero and variance $\eta_k$. In the analysis of a sample of curves, the expansion in (\ref{eq:fpca_conceptual}) is truncated to only retain the first $K$ eigenfunctions. All terms in the truncated KKL expansion can be estimated directly using a probabilistic- or likelihood-based approach as an alternative to decomposing of a covariance operator \citep{tipping_probabilistic_1999,james2000principal,van2009bayesian,goldsmith2015generalized}. 

In real data settings, we observe $Y_i(t) = X_i(t) + \epsilon_i(t)$, where $\epsilon_i(t)$ is assumed to be white noise with fixed variance. Functions are additionally observed over a discrete grid of timepoints $\boldsymbol{t}_i = \Big\{t_{ij}: j \in \{1,\dots,J_i\}\Big\}$ which may vary across subjects. Let $Y_i(t_{ij})$ be the value of $Y_i(\cdot)$ evaluated at $t_{ij}$, and $Y_i(\boldsymbol{t}_i)$ be the $J_i \times 1$ vector of $Y_i(\cdot)$ evaluated over $\boldsymbol{t}_i$; similar notation will be used for other functions so that, for example, $\mu(\boldsymbol{t}_i)$ is the vector containing the mean $\mu(\cdot)$ evaluated over $\boldsymbol{t}_i$. We pursue a spline basis approach to fitting (\ref{eq:fpca_conceptual}), and express the mean $\mu(t)$ and eigenfunctions $\mathbf{\Phi}(t)$ using a spline basis  $\mathbf{W}(t) = \Big\{w_p(t): p \in \{1,\dots, P\}\Big\}$. Let $\mathbf{W}(t_{ij})$ be the $1 \times P$ vector containing the spline basis evaluated at $t_{ij}$ and $\mathbf{W}(\boldsymbol{t}_i)$ be the $J_i \times P$ matrix containing the spline basis evaluated over the vector $\boldsymbol{t}_i$; $\boldsymbol{\beta}_{\mu}$ be a $P \times 1$ vector of spline coefficients corresponding to $\mu(t)$; and $\boldsymbol{\beta}_{\boldsymbol{\Phi}} = \left[\boldsymbol{\beta}_{\phi_1},\dots, \boldsymbol{\beta}_{\phi_K} \right]$ be the $P \times K$ matrix of spline coefficients corresponding to $\boldsymbol{\Phi}(t)$. Using these, we define the spline expansions $\mu(\boldsymbol{t}_i)=\mathbf{W}(\boldsymbol{t}_i)\boldsymbol{\beta}_{\mu}$ and $\boldsymbol{\Phi}(\boldsymbol{t}_i) = \left[\phi_1(\boldsymbol{t}_i), ...,\phi_K(\boldsymbol{t}_i)\right]^T = \mathbf{W}(\boldsymbol{t}_i) \boldsymbol{\beta}_{\boldsymbol{\Phi}}$. Lastly, let $\boldsymbol{\xi}_i = \Big\{\xi_{ik}: k \in \{1,\dots,K\}\Big\}$ be the $K \times 1$ vector of scores for observation $i$. In this manuscript, we will use $\mathbf{W}(t)$ to denote a spline basis with orthonormal second derivatives, used in implementations of smoothing splines such as \texttt{mcgv::gam()} \citep{wood2006low,wood2017generalized}. 

With the preceding notation, we recast (\ref{eq:fpca_conceptual}) for observed data $Y_i(\cdot)$ measured at timepoint $t_{ij}$ by

\begin{equation}
\label{eq:fpca_probabilistic}
\begin{split}
  Y_i(t_{ij})= \mu(t_{ij}) + \sum_{k=1}^{K}\xi_{ik}\phi_k(t_{ij}) + \epsilon_{i}(t_{ij}) \\ 
   = \mathbf{W}(t_{ij})\boldsymbol{\beta}_\mu + \mathbf{W}(t_{ij})\boldsymbol{\beta}_{\mathbf{\Phi}}\xi_i + \epsilon_{i}(t_{ij}) \\
\end{split}
\end{equation}

\noindent
where $\epsilon_{i}(t_{ij})$ is noise with unknown variance $\sigma^2_\epsilon$. Making the common distributional assumptions that $\epsilon_{i}(t_{ij}) \sim \text{N}\left(0, \sigma^2_\epsilon\right)$ and $\boldsymbol{\xi}_{i} \sim \text{MVN}\left(0,\mathbf{I}_{K \times K}\right)$, where $\mathbf{I}_{K \times K}$ is the $K \times K$ identity matrix, it is possible to estimate the coefficients of (\ref{eq:fpca_probabilistic}) by finding the maximum likelihood estimates of the spline coefficients $\boldsymbol{\beta}_\mu$ and $\boldsymbol{\beta}_{\mathbf{\Phi}}$, the scores $\boldsymbol{\Xi} = \Big\{\boldsymbol{\xi}_i: i \in \{1,\dots, I\}\Big\}$, and the error variance $\sigma^2_\epsilon$. In practice, including standard second derivative penalties to enforce smoothness of the mean and FPCs is common, as is a post-processing step to ensure FPCs are orthonormal.

\subsection{Smoothing via an Adaptive Ridge Penalty}
\label{subsec:adaptivesmooth_scatterplot}

The approach to adaptive scatterplot smoothing we develop in this subsection is a central contribution of this manuscript, and later will be used in the context of FPCA. To the extent possible, we retain notation introduced in the previous section. Assume we observe data $\Big\{ (t_j,y_j) : j \in \{1,\dots,J\} \Big\}$ and that $y_j \sim N\left[f(t_j), \sigma^2_{\epsilon}\right]$. The goal of scatterplot smoothing is to flexibly estimate the unknown function $f(\cdot)$ defined over $t \in \mathcal{T}$. A spline-based estimator of $f(t)$ can be obtained by expanding $f(t_j) = \mathbf{S}(t_j) \boldsymbol{\alpha}_f$ using the basis $\mathbf{S}(t)$ and the vector of coefficients $\boldsymbol{\alpha}_f$, and maximizing the likelihood with respect to $\boldsymbol{\alpha}_f$. 

Spline-based methods can explicitly protect against overfitting by imposing a penalty on the complexity of the estimate $\hat{f}(t)$; the squared-second-derivative penalty is a common choice. Define $\mathbf{S}^{''}(t) = \Big\{s_p^{''}(t): p \in \{1,\dots, P\}\Big\}$ to be the second derivatives of the spline functions $\mathbf{S}(t)$. Note that, in contrast to the basis $\mathbf{W}(t)$ in the previous section, we do not require $\mathbf{S}(t)$ to have orthonormal second derivatives; we choose $\mathbf{S}(t)$ to be a cubic B-spline basis. Then $\lambda \int_{\mathcal{T}} \left[f^{\prime \prime}(t) \right]^2 dt =  \lambda \sum_{p=1}^P \sum_{q=1}^P \left[\alpha_{fp} \alpha_{fq} \int_{\mathcal{T}} s_p^{\prime \prime}(t) s_q^{\prime \prime}(t)dt\right]$ is the squared-second derivative penalty, and $\lambda$ is the associated tuning parameter. More compactly, let $\boldsymbol{\Omega}_{\mathbf{S}}$ be the $P \times P$ penalty matrix with entries 
\begin{equation*}
\label{eq:omega}
 {\boldsymbol{\Omega}_{\mathbf{S}}}_{pq}=\int_{\mathcal{T}} s_{p}^{\prime \prime}(t) s_{q}^{\prime \prime}(t) d t, \quad p,q \in \{1, \dots, P\}
\end{equation*}

\noindent
so that $\int_{\mathcal{T}} \left[f^{\prime \prime}(t)\right]^2 dt = \boldsymbol{\alpha}_f^T \boldsymbol{\Omega}_{\mathbf{S}} \boldsymbol{\alpha}_f$. Setting $\mathbf{y} = \Big\{y_j: j \in \{1,\dots,J\}\Big\}$ and $\boldsymbol{t} = \Big\{t_j: j \in \{1,\dots,J\}\Big\}$ to be $J \times 1$ observation vectors, and $\mathbf{S}(\boldsymbol{t})$ to be a $J \times P$ matrix containing values of $\mathbf{S}(t)$ evaluated over $\boldsymbol{t}$, the penalized likelihood used to estimate $f(\cdot)$ in a non-adaptive fashion is

\begin{equation*}
  \label{eq:penalized_likelihood}
    \ell(\boldsymbol{\alpha}_f,\sigma_\epsilon^2; \boldsymbol{y}, \boldsymbol{t}, \boldsymbol{\Omega}_{\mathbf{S}}) = -\frac{I}{2}\log(\sigma^2_\epsilon) -\frac{I}{2\sigma^2_\epsilon} \left\|\mathbf{y} - \mathbf{S}(\boldsymbol{t}) \boldsymbol{\alpha}_f \right\|^2 + \lambda \boldsymbol{\alpha}_f^T \boldsymbol{\Omega}_{\mathbf{S}} \boldsymbol{\alpha}_f.
\end{equation*}

\noindent
A mathematically equivalent likelihood can be obtained by treating spline coefficients $\boldsymbol{\alpha}$ as random effects with a covariance equal to the (generalized) inverse $\boldsymbol{\Omega}_{\mathbf{S}}^{+}$. Doing so relates the tuning parameter to the variance of the residuals and random effects, so that the tuning parameter can be estimated from data rather than using a computationally expensive cross validation procedure.

We achieve adaptive smoothing by replacing the tuning parameter $\lambda$ by a tuning function $\lambda(t)$ defined over $\mathcal{T}$, and implementing the tuning function through an adaptive ridge penalty. Let $\lambda(t) = \left[\sum_{p=1}^P \lambda_p m_p(t)\right]^2$ where $\mathbf{M}(t) = \Big\{m_p(t): p \in \{1,\dots, P\}\Big\}$ is a spline expansion of the same dimension as $\mathbf{S}(t)$ and $\boldsymbol{\lambda} = \Big\{\lambda_p: p \in \{1,\dots, P\}\Big\}$ is the corresponding $P \times 1$ vector of coefficients. The quadratic form ensures the required constrain that  $\lambda(t) \geq 0$ for all $t \in \mathcal{T}$. Using the spline bases $\mathbf{S}(t)$ and $\mathbf{M}(t)$ with coefficients $\boldsymbol{\alpha}_f$ and $\boldsymbol{\lambda}$ to express $f(t)$ and $\lambda(t)$, respectively, the adaptive smoothing penalty is given by:
\begin{equation}
  \label{eq:adaptive_penalty_expanded}
    \begin{split}
    \int_{\mathcal{T}} \lambda(t) f^{\prime \prime}(t)^2 dt &= 
    \sum_{p=1}^P \sum_{q=1}^P \sum_{u=1}^P \sum_{v=1}^P \left[ \lambda_p \lambda_q \alpha_{fu} \alpha_{fv} \int_{\mathcal{T}} m_p(t) m_q(t) s_u^{\prime \prime}(t) s_v^{\prime \prime}(t)dt \right]
    \end{split}.
\end{equation}

While (\ref{eq:adaptive_penalty_expanded}) holds for any choice of $\mathbf{S}(t)$ and $\mathbf{M}(t)$, we will construct $\mathbf{W}(t)$ and $\mathbf{M}(t)$ from $\mathbf{S}(t)$ such that $\mathbf{W}(\boldsymbol{t})\boldsymbol{\beta}_f = \mathbf{S}(\boldsymbol{t}) \boldsymbol{\alpha}_f$ and

\begin{equation}
  \label{eq:adaptive_penalty}
\int_{\mathcal{T}} \lambda(t) f^{\prime \prime}(t)^2 dt = \sum_{p=1}^P \lambda_p^2 {\beta_{fp}^2} = \boldsymbol{\beta}_f^T \boldsymbol{\Lambda} \boldsymbol{\beta}_f
\end{equation}

\noindent
where $\boldsymbol{\beta}_f$ are coefficients and $\boldsymbol{\Lambda} = \text{diag}\!\left\{\lambda_1^2, \dots, \lambda_p^2\right\}$ is a $P \times P$ diagonal matrix. That is, the adaptive smoothing penalty (\ref{eq:adaptive_penalty_expanded}) is expressed as an adaptive ridge penalty in which the spline coefficients $\boldsymbol{\beta}_f$ for the function $f(\cdot)$ are weighted by the spline coefficients $\boldsymbol{\lambda}$ for $\lambda(\cdot)$. The procedure to construct $\mathbf{W}(t)$ and $\mathbf{M}(t)$ such that (\ref{eq:adaptive_penalty}) holds will be presented shortly; first we describe our estimation approach assuming these bases are available. 

Define $\mathbf{W}(\boldsymbol{t})$ to be $J \times P$ matrix of values of $\mathbf{W}(t)$ evaluated at $\boldsymbol{t}$. We propose to estimate $f(\cdot)$ by maximizing the penalized likelihood

\begin{equation}
  \label{eq:penalized_likelihood_neww}
    \ell(\boldsymbol{\beta}_f,\sigma_\epsilon^2,\boldsymbol{\Lambda} ; \boldsymbol{y}, \boldsymbol{t}) = -\frac{I}{2}\log(\sigma^2_\epsilon) -\frac{I}{2\sigma^2_\epsilon} \left\|\mathbf{y} - \mathbf{W}(\boldsymbol{t}) \boldsymbol{\beta}_f \right\|^2 + {\boldsymbol{\beta}_f}^T \boldsymbol{\Lambda} \boldsymbol{\beta}_f
\end{equation}

\noindent
where $\boldsymbol{\Lambda}$ is the diagonal matrix containing unique tuning parameters for the elements of $\boldsymbol{\beta}_f$. We maximize (\ref{eq:penalized_likelihood_neww}) with an algorithm that iterates between updating coefficients $\boldsymbol{\beta}_f$, tuning parameters $\boldsymbol{\Lambda}$, and the residual variance $\sigma_\epsilon^2$ using the following estimators:

\begin{itemize}

  \item $\boldsymbol{\hat{\beta}}_f = \left[ {\mathbf{W}(\boldsymbol{t})}^{T} \mathbf{W}(\boldsymbol{t})  + \sigma_\epsilon^2 \boldsymbol{\Lambda} \right]^{-1} \left[{\mathbf{W}(\boldsymbol{t})}^T \mathbf{y}\right]$
  
  \item $\hat{\boldsymbol{\Lambda}} = \text{diag}\!\left\{0,0,\hat{\lambda}_3^2, \dots, \hat{\lambda}_P^2\right\} \quad \text{where} \quad \hat{\lambda}_p = \frac{1}{{\boldsymbol{\beta}_f}_{p}}$

  \item $\hat{\sigma}_\epsilon^2 = \|\mathbf{y} - \mathbf{W}(\boldsymbol{t}) \boldsymbol{\beta}_f \|^2 / J$.
  
\end{itemize}

We initialize our algorithm by letting ${\boldsymbol{\Lambda}}^{(0)} = \boldsymbol{0}_{P \times P}$ and obtaining unpenalized coefficients ${\boldsymbol{\beta}}^{(0)}_f = \left[ \mathbf{W}(\boldsymbol{t})^{T} \mathbf{W}(\boldsymbol{t})\right]^{-1}\Big[\mathbf{W}(\boldsymbol{t})^T \mathbf{y}\Big]$. At each iteration, we evaluate (\ref{eq:penalized_likelihood_neww}) given the current parameter estimates, and monitor convergence using the absolute difference between the current evaluated penalized likelihood and its previous estimate. Both $\lambda_1$ and $\lambda_2$ are zero due to the inclusion of two unpenalized basis functions in our $\mathbf{W}(\boldsymbol{t})$; this point is further elaborated in the following paragraph. Our approach builds on the perspective of smoothing splines as mixed effects models, in that our estimated tuning parameters correspond to solutions of a random-effects model in which each coefficient has an independent Gaussian prior with a coefficient-specificvariance. In practice, we include a lower limit of $10^{-6}$ for the components of $\boldsymbol{\beta}_f$ to prevent arithmetic overflow, as tuning parameters tend toward infinity for coefficients that approach zero.

 We now discuss our strategy for constructing bases $\mathbf{W}(t)$ and $\mathbf{M}(t)$ that satisfy (\ref{eq:adaptive_penalty}). Our approach follows \citet{wand_semiparametric_2008}, who used a similar transformation to obtain a simple mixed model representation for non-adaptive smoothing. We begin with $\mathbf{S}(t)$ and the corresponding second-derivative penalty matrix $\boldsymbol{\Omega}_\mathbf{S}$. An eigendecomposition of $\boldsymbol{\Omega}_\mathbf{S}$ yields an orthogonal matrix $\mathbf{Q}$ and a diagonal matrix $\boldsymbol{\Psi}$ containing the eigenvectors and eigenvalues of $\boldsymbol{\Omega}_\mathbf{S}$, respectively, so that $\boldsymbol{\Omega}_\mathbf{S} = \mathbf{Q}^T\boldsymbol{\Psi}\mathbf{Q}$. 
When $\mathbf{S}(\boldsymbol{t})$ is a cubic B-spline basis, it will span the space of straight lines, but the penalty is on second derivatives. As a result, $\text{rank}(\boldsymbol{\Omega}_\mathbf{S}) = P - 2$ and $\boldsymbol{\Psi}$ will have two zero-entries and $P-2$ positive entries \citep{speed1991blup}. Define partitions $\mathbf{Q}= \left[\mathbf{Q}_1 | \mathbf{Q}_2 \right]$ and 
$\boldsymbol{\Psi} = \left[\boldsymbol{\Psi}_1 | \boldsymbol{\Psi}_2 \right]$ such that $\mathbf{Q}_2$ and $\boldsymbol{\Psi}_2$ are the sub-matrices of $\mathbf{Q}$ and $\boldsymbol{\Psi}$ with columns that correspond to the non-zero eigenvalues in $\boldsymbol{\Psi}$. Finally, let $\mathbf{U} = \left[ \mathbf{Q}_1 | \mathbf{Q}_2 \boldsymbol{\Psi}_2^{-1/2} \right]$ and define ${\mathbf{W}}(\boldsymbol{t}) = \mathbf{S} (\boldsymbol{t}) \mathbf{U}$. The second derivatives of the transformed basis ${\mathbf{W}}(t)$ satisfy

 \begin{equation*}
  \label{eq:omega_transformation}
    {\mathbf{W}}^{\prime \prime} (\boldsymbol{t})^T {\mathbf{W}}^{\prime \prime}(\boldsymbol{t})= 
    \begin{bmatrix}
    \boldsymbol{0}_{2 \times 2} & \boldsymbol{0}\\
    \boldsymbol{0} & \mathbf{I}_{(P-2) \times (P-2)}
    \end{bmatrix}.
\end{equation*}
 
\noindent
We therefore set $\lambda_1=\lambda_2=0$ and let $w_1(t)$ and $w_2(t)$ be the (unpenalized) intercept and slope basis functions. Choosing $\mathbf{M}(\boldsymbol{t})$ so that 

 \begin{equation*}
  \label{eq:m_basis}
\mathbf{M}(\boldsymbol{t}) = \left\{m_q(\boldsymbol{t}) = \frac{{\beta_f}_q {w_q}^{\prime \prime}(\boldsymbol{t})}{\sum_{p = 3}^P {\beta_f}_p {w_p}^{\prime \prime}(\boldsymbol{t})}: \linebreak q \in \{3,\dots, P\}\right\}
\end{equation*}

\noindent
will ensure (\ref{eq:adaptive_penalty}) holds; as a consequence, adaptive scatterplot smoothing can be represented as the likelihood that includes an adaptive ridge penalty given in (\ref{eq:penalized_likelihood_neww}) . 

The basis $\mathbf{M}(t)$ does not affect the estimation of $\boldsymbol{\beta}_f$ or $\boldsymbol{\Lambda}$, and it is sufficient to know that such a basis exists for the algorithm given above to be well-defined. Moreover, since $\mathbf{M}(t)$ depends on the current estimate of $\boldsymbol{\beta}_f$, this basis varies across iterations (although $\mathbf{W}(t)$ remains fixed). At any iteration, however, the current estimate of the penalty function can be obtained as $\hat{\lambda}(t) = \left\{\frac{\sum_{p=1}^P \left[\hat{\lambda}_p {\hat{\beta}_{f_p}} {w_p}^{\prime \prime}(t)\right]}{\sum_{p=1}^P \left[{\hat{\beta}}_{f_p} {w_p}^{\prime \prime}(t)\right]}\right\}^2$. Under this framework, each of the penalized coefficients and their corresponding spline basis carries different local and global information about the overall smoothness of the estimated fit, and allowing each of these splines to be weighted differently relaxes the assumption that the smoothness across the fit is equal.

\subsection{Adaptive Smoothing Functional Principal Component Analysis}
\label{subsec:our_method}

Our primary objective in this manuscript is to estimate patterns of variation shared across functional observations using FPCA; the latent functions are also used to reconstruct and denoise individual curves. We smooth each FPC adaptively to capture local differences in smoothness across the functional domain. We accomplish this by extending the FPCA framework in Section \ref{subsec:fpca_model} to estimate the mean function $\mu(t)$ and set of FPCs $\boldsymbol{\Phi}(t) = \Big\{\phi_k(t): k \in \{1,\dots,K\}\Big\}$ using the the adaptive smoothness penalty developed for scatterplot smoothing in Section \ref{subsec:adaptivesmooth_scatterplot}. 

Express $\mu(t_i)=\mathbf{W}(\boldsymbol{t}_i)\boldsymbol{\beta}_{\mu}$  and $\boldsymbol{\Phi}(\boldsymbol{t}_i) = \left[\phi_1(\boldsymbol{t}_i), ...,\phi_K(\boldsymbol{t}_i)\right]^T = \mathbf{W}(\boldsymbol{t}_i) \boldsymbol{\beta}_{\boldsymbol{\Phi}}$ using the transformed cubic B-spline basis $\mathbf{W}(t)$ such that condition (\ref{eq:adaptive_penalty}) holds,  with corresponding coefficients $\boldsymbol{\beta}_\mu$ and  $\boldsymbol{\beta}_{\boldsymbol{\Phi}}$. We maximize 
\begin{equation}
\label{eq:adap_penalized_likelihood}
\begin{split}
  & \ell\left(\boldsymbol{\beta}_\mu, \boldsymbol{\beta}_{\boldsymbol{\Phi}}, \sigma^2_\epsilon ; \bY,\boldsymbol{\Xi}\right) - P = \\
  & \ell\left(\boldsymbol{\beta}_\mu, \boldsymbol{\beta}_{\boldsymbol{\Phi}}, \sigma^2_\epsilon ; \bY| \boldsymbol{\Xi}\right)  + \ell(\boldsymbol{\Xi}) - P \propto \\
  & \sum_{i = 1}^I \left(\frac{\|Y_i(\boldsymbol{t}_i) - \mathbf{W}(\boldsymbol{t}_i)\boldsymbol{\beta}_{\mu} - \mathbf{W}(\boldsymbol{t}_i)\boldsymbol{ \beta}_{\boldsymbol{\Phi}} \boldsymbol{\xi}_i\|^2}{2\sigma^2_\epsilon}\right) +
  \sum_{i = 1}^I \left(\frac{\left\|\boldsymbol{\xi}_{i}\right\|^2}{2}\right) 
   + \sum_{i=1}^I \left(\frac{J_i \log\sigma^2_\epsilon }{2}\right)
   - P
  \end{split}
\end{equation}
\noindent
with respect to $\boldsymbol{\beta}_\mu$ and $\boldsymbol{\beta}_{\boldsymbol{\Phi}}$, $\sigma^2_\epsilon$ and $\boldsymbol{\Xi}$. The penalty term $P(\boldsymbol{\beta}_\mu, \boldsymbol{\beta}_{\boldsymbol{\Phi}})$ is defined as: 
\begin{equation*}
\label{eq:adap_penalty}
\begin{split}
  P &= \int_{\mathcal{T}} \left[\lambda_{\mu}(t) \mu^{\prime \prime}(t)^2 dt\right] + \sum_{k=1}^K \left\{ \int_{\mathcal{T}} \left[ \lambda_{\phi_k}(t) \phi_k^{\prime \prime}(t)^2 dt\right] \right\}
  \end{split}.
\end{equation*}

\noindent
Using the techniques described in Section \ref{subsec:adaptivesmooth_scatterplot} this penalty can be expressed as
\begin{equation*}
\label{eq:adap_penalty_ridge}
\begin{split}
    P = {\boldsymbol{\beta}_\mu^T} \boldsymbol{\Lambda_\mu} \boldsymbol{\beta}_\mu + 
       \sum_{k =1}^K {\boldsymbol{\beta}_{\phi_k}^T} \boldsymbol{\Lambda}_{\phi_k} \boldsymbol{\beta}_{\phi_k}
  \end{split}
\end{equation*}

\noindent
where $\boldsymbol{\Lambda}_\mu$ and $\boldsymbol{\Lambda}_{\phi_k}$, $k = 1,\dots,K$, are diagonal $ P \times P$ matrices of tuning parameters (e.g. $\boldsymbol{\Lambda}_\mu = \text{diag}\!\left\{0,0,{\lambda_\mu}_3,\dots, {\lambda_\mu}_P\right\}$). As in Section \ref{subsec:adaptivesmooth_scatterplot}, this provides direct estimates of the tuning functions $\lambda_{\mu}(t)$ and $\lambda_{\phi_k}(t)$, $k = 1,\dots,K$, as part of the adaptive FPCA approach. 


\subsubsection{Algorithm and Implementation}
\label{subsubsec:algorithm}

The estimation algorithm for adaptive FPCA iterates between the following steps:
\begin{enumerate}[itemsep=-2mm]

  \item The mean $\mu (t)$ and FPCs $\mathbf{\Phi}(t)$ are estimated  by maximizing (\ref{eq:adap_penalized_likelihood}) with respect to $\boldsymbol{\beta}_\mu$ and $\boldsymbol{\beta}_{\boldsymbol{\Phi}}$, conditional on the current estimates of scores $\boldsymbol{\Xi}$, tuning parameters $\boldsymbol{\Lambda}_\mu$ and $\boldsymbol{\Lambda}_{\phi_k}$, $k = 1,\dots,K$, and $\sigma_\epsilon^2$. 

  \item Subject-specific scores $\boldsymbol{\xi}_i$ are estimated given current estimates of coefficients $\boldsymbol{\beta}_\mu$ and $\boldsymbol{\beta}_{\boldsymbol{\Phi}}$, and of the residual variance $\sigma_\epsilon^2$.

  \item Tuning parameters $\boldsymbol{\Lambda}_\mu$ and $\boldsymbol{\Lambda}_{\phi_k}$, $k = 1,\dots,K$, are estimated given current values of  $\boldsymbol{\beta}_\mu$ and $\boldsymbol{\beta}_{\boldsymbol{\Phi}}$. The residual variance $\sigma_\epsilon^2$ is estimated using the method of moments.
  
\end{enumerate}

Each step of the algorithm and the initialization is described in detail below. Following each iteration, we evaluate (\ref{eq:adap_penalized_likelihood}) given the current parameter estimates, and monitor the convergence using the absolute difference between the current evaluated penalized likelihood and its previous estimate.

\subsubsection*{Spline Coefficient Estimation}
\label{subsubsec:est_fpc}

Define $\boldsymbol{Y} = \Big[Y_1(\boldsymbol{t}_1),...,Y_I(\boldsymbol{t}_I)\Big]^T$ to be the $\left(\sum_{i=1}^I J_i \right) \times 1$ vector of concatenated observations $Y_i(\boldsymbol{t}_i)$. For each subject $i$, denote $\boldsymbol{\theta}_i = \left(1, \boldsymbol{\xi}_i\right)$ to be a vector of dimension $1 \times (K+1)$ and define the matrix $\boldsymbol{\Theta}_i(\boldsymbol{t}_i) = \boldsymbol{\theta}_i \otimes \mathbf{W}(\boldsymbol{t}_i)$ where $\otimes$ is the Kronecker product. Further, let $\boldsymbol{B} = \Big[{\boldsymbol{\beta}^{T}_\mu}, {\boldsymbol{\beta}^{T}_{\phi_1}}, \dots, {\boldsymbol{\beta}^{T}_{\phi_K}}\Big]^T$ be a $(K+1)P \times 1$ vector of concatenated coefficient vectors. Using this notation, (\ref{eq:fpca_probabilistic}) can be rewritten as $\boldsymbol{Y} = \mathbf{\Theta} \boldsymbol{B} + \boldsymbol{E}$ where $\boldsymbol{E}$ is the $\left(\sum_{i=1}^I{J_i} \right) \times 1$ vector of concatenated $\epsilon_i(\boldsymbol{t}_j)$ and $\mathbf{\Theta} =  \Big[\boldsymbol{\Theta}_1(\boldsymbol{t}_1)^T,\dots,\boldsymbol{\Theta}_I(\boldsymbol{t}_I)^T\Big]^T$. Define $\boldsymbol{\Lambda}_{\mu} = \text{diag}\!
\left\{ 0,0,\lambda_{\mu3}, \dots, \lambda_{\mu p} \right\}$ and $\boldsymbol{\Lambda}_{\phi_k} = \text{diag}\! \left\{ 0,0, \lambda_{\phi_k 3}, \dots, \lambda_{\phi_k p} \right\}$ for each $k \in 1,\dots,K$ to be a $P \times P$ diagonal matrix of tuning parameters, and $\mathbf{0}_{P \times P}$ to be the zero matrix. The estimate for $\boldsymbol{B}$ is simply

\begin{equation*}
\label{eq:coefficient_estimate}
\boldsymbol{\hat{B}} = \Big(\mathbf{\Theta}^T \mathbf{\Theta}  + \sigma^2_\epsilon \boldsymbol{\Lambda}\Big)^{-1} \Big(\mathbf{\Theta}^{T}  \boldsymbol{Y} \Big) 
\end{equation*}

\noindent 
where $\boldsymbol{\Lambda}$ is defined as

\begin{equation*}
  \label{eq:Lambda_FPCA_All}
    \boldsymbol{\Lambda} = 
  \begin{pmatrix}
    \boldsymbol{\Lambda}_{\mu} & \mathbf{0}_{P \times P } & \dots & \mathbf{0}_{P \times P } \\
    \mathbf{0}_{P \times P } & \boldsymbol{\Lambda}_{\phi_1} & \dots & \mathbf{0}_{P \times P } \\
    \vdots & \vdots & \ddots & \vdots \\
    \mathbf{0}_{P \times P } & \mathbf{0}_{P \times P } & \dots & \boldsymbol{\Lambda}_{\phi_K}
  \end{pmatrix}.
\end{equation*}

\noindent
As in adaptive and non-adaptive forms of scatterplot smoothing, the estimated spline coefficients have closed form solutions given current values of the tuning parameters.

\subsubsection*{Score Estimation}
\label{subsubsec:est_scores}

At each iteration, the subject-specific scores are calculated by maximizing the full likelihood (\ref{eq:adap_penalized_likelihood}) given all other parameters. For subject $i$ with observed values $Y_i(\boldsymbol{t}_i)$, the estimate of $\boldsymbol{\xi}_i$ is:

\begin{equation*}
\label{eq:scores_estimate}
\boldsymbol{\hat{\xi}}_{i} = \left[\frac {\boldsymbol{\Phi}(\boldsymbol{t}_i)^T \boldsymbol{\Phi}(\boldsymbol{t}_i)}{\sigma^2_\epsilon} + \mathbf{I}_{K \times K}\right]^{-1} \bigg\{\boldsymbol{\Phi}(\boldsymbol{t}_i)^T \big[Y_i(\boldsymbol{t}_i) - \mu(\boldsymbol{t}_i)\big]\bigg\}.
\end{equation*}

\noindent
That is, at each iteration we estimate scores using BLUPs with current values of other parameters, following common practice in FPCA. 

\subsubsection*{Adaptive Smoothness Tuning Parameters}
\label{subsubsec:est_adaptive}

We update the adaptive smoothing weights given the present estimates of the coefficients using the same approach as in adaptive scatterplot algorithm in Section \ref{subsec:adaptivesmooth_scatterplot}. Given the coefficients $\boldsymbol{B} = \Big[{\boldsymbol{\beta}^{T}_\mu}, {\boldsymbol{\beta}^{T}_{\phi_1}}, \dots, {\boldsymbol{\beta}^{T}_{\phi_K}}\Big]^T$, we have

\begin{equation*}
\label{eq:lambda}
\begin{split}
\hat{\boldsymbol{\Lambda}}_\mu &= \text{diag}\!\left\{0,0, \left(\frac{1}{\beta_{\mu3 }}\right)^2, \dots, \left(\frac{1}{\beta_{\mu P}}\right)^2\right\} \\
\hat{\boldsymbol{\Lambda}}_{\phi_k} &= \text{diag}\!\left\{0,0, \left(\frac{1}{\beta_{{\phi_k}3}}\right)^2, \dots, \left(\frac{1}{\beta_{{\phi_k}P}}\right)^2\right\}, k \in \{1,\dots,K\}
\end{split}
\end{equation*}

\noindent
as the updated estimates of the respective tuning parameters for each set of coefficients. The residual variance $\sigma^2_\epsilon$ is estimated as 

\begin{equation*}
\label{eq:sigma_hat}
\hat{\sigma}^2_\epsilon = \left[\frac{\sum_{i = 1}^I \|Y_i(\boldsymbol{t}_i) - \mathbf{W}(\boldsymbol{t}_i)\boldsymbol{\beta}_{\mu} - \mathbf{W}(\boldsymbol{t}_i)\boldsymbol{ \beta}_{\boldsymbol{\Phi}} \boldsymbol{\xi}_i\|^2}{\sum_{i = 1}^I \left(J_i\right)}\right]. 
\end{equation*}

\subsubsection{Practical Concerns}
\label{subsubsec:practical_concerns}

The iterative algorithm can be initialized using random values for $\boldsymbol{\Xi}^{(0)}$ where $\boldsymbol{\xi}_i^{(0)} \sim N(0, \mathbf{I}_{K \times K})$, but convergence is faster if reasonable starting values are provided. We use a computationally efficient non-adaptive FPCA method, FACE, implemented in the \texttt{refund} package \citep{xiao2018fast,goldsmith2016refund}. After initializing $\boldsymbol{\Xi}$, we set tuning parameters as $\boldsymbol{\Lambda}_\mu^{(0)} = \boldsymbol{\Lambda}_{\phi_k}^{(0)} = \mathbf{0}_{P \times P} \quad \forall k \in 1,\dots,K$, and find initial estimates of $\boldsymbol{\beta}_{\mu}$, $\boldsymbol{\beta}_{\Phi}$, and $\sigma_\epsilon^2$ using ordinary least squares.

Likelihood-based approaches to FPCA differ from techniques that decompose a covariance in some notable but addressable ways. First, like many probabilistic approaches to PCA and FPCA, we do not constrain the FPCs to be orthogonal in the estimation of spline coefficients. Instead,  orthogonalize estimated FPCs using a rotation step in each iteration. Second, it is necessary to pre-specify the number of FPCs estimated in the model. We begin by modeling a large enough number of FPCs to capture the explainable variability in the data  and then truncate to retain only the components that explain a large percent of the variance. We typically set K = 15 and the percent variance explained to 99\%, although we recommend sensitivity analyses to these choices. 

Our proposed adaptive scatterplot smoothing and adaptive FPCA have publicly available software implementations in the R package $\texttt{afpca}$. While our manuscript assumes that an initial cubic B-spline basis $\mathbf{S}(t)$, our software has been implemented for a variety of spline basis. Our user interface accepts several data structures, including recent $\texttt{tidyfun}$ tools for tidy functional data analysis \citep{tidyfun}. The package is available at \url{https://github.com/angelgar/afpca}.


\section{Simulations}
\label{sec:sim} 

We illustrate the performance of our proposed method  using simulated data that mimics our motivating study. In particular, we examine the ability of our adaptive FPCA to estimate FPCs and reconstruct individual curves, and compare to existing non-adaptive FPCA methods. 

\subsection{Simulation Design}
\label{subsec:sim_design}

We generate curves $Y_i(t)$ according to the FPCA model

\begin{equation*}
\label{eq:fpca_observed}
  Y_i(t) = \mu(t) + \sum_{k=1}^{2}\xi_{ik}\phi_k(t) + \epsilon_i(t)
\end{equation*}

\noindent
over an equally spaced grid of 100 timepoints, $\boldsymbol{t} \in [0,1]$, shared across all observations $ 1 \leq i \leq I$. The mean and FPCs are defined in a piecewise fashion, and exhibit varying degrees of smoothness over the functional domain. Specifically, we let

\begin{equation*}
\label{eq:simulation_formula}
    \begin{split}
    \mu(t)  &= c_\mu  t^{-\frac{3}{2}} \sin\!\left(\pi  t^{\frac{1}{4}}\right) \mathcal{I}\left[t >  \frac{1}{2}\right]   \\
    \phi_k(t) &=  c_{\phi_k}  t^{-\frac{3}{2}} \sin\!\left(4k \pi  t^{\frac{1}{4}}\right)  \mathcal{I}\left[t >  \frac{1}{2}\right], \: k \in \{1,2\}
    \end{split}
\end{equation*}

\noindent
where $c_{\mu} = \int_{\frac{1}{2}}^1 t^{-\frac{3}{2}} \sin\!\left(\pi  t^{\frac{1}{4}}\right) dt$ and $c_{\phi_k} = \int_{\frac{1}{2}}^1 t^{-\frac{3}{2}} \sin\!\left(4k \pi  t^{\frac{1}{4}}\right)dt$, $k = \{1,2\}$ are normalizing constants, and $\mathcal{I}[\cdot]$ is the indicator function. The mean and FPCs are constant in the first half of the observation window and are sine functions with a smoothly changing period over the second half of the observation window, and the FPCs are orthonormal. Scores $\xi_{ik}$ are generated from a mean-zero Normal distribution with $\eta^2_1  = 4$ and $\eta^2_2  = 1$. Finally, we draw residuals $\epsilon_i(t_{ij})$ from a mean-zero Normal distribution with variance $\sigma_\epsilon^2$. 

We generate 100 datasets for each combination of sample sizes $I \in \{25, 50, 100\}$ and error variances $\sigma^2_\epsilon \in \{0.1, 0.2\}$, and apply our adaptive FPCA method to each simulated dataset. We compare it to two functions in the R package \texttt{refund} \citep{goldsmith2016refund}: \texttt{fpca.sc()} and \texttt{fpca.face()}. \texttt{fpca.sc()} uses penalized tensor product of B-splines to smooth the estimated covariance operator before eigendecomposition, while \texttt{fpca.face()} is a faster method for smoothing the covariance \citep{xiao2018fast}. We compare estimation accuracy using the (mean) integrated square error (MISE), taken as the average of $ISE = \int_0^1 \left[ Y_i(t) - \hat{Y}_i(t) \right]^2 dt$ across subjects. We also compute the ISEs for the mean and FPCs. For all implementations, we fix the number of spline basis functions at $P=40$, determine the number of retained FPCs using a 99\% variance explained threshold, and report retained FPCs estimated by each method. 

\subsection{Simulation Results}
\label{subsec:sim_results}

\begin{figure}[htb]
  \centering
     \begin{tabular}{cc}
       \includegraphics[width= 0.8\textwidth]{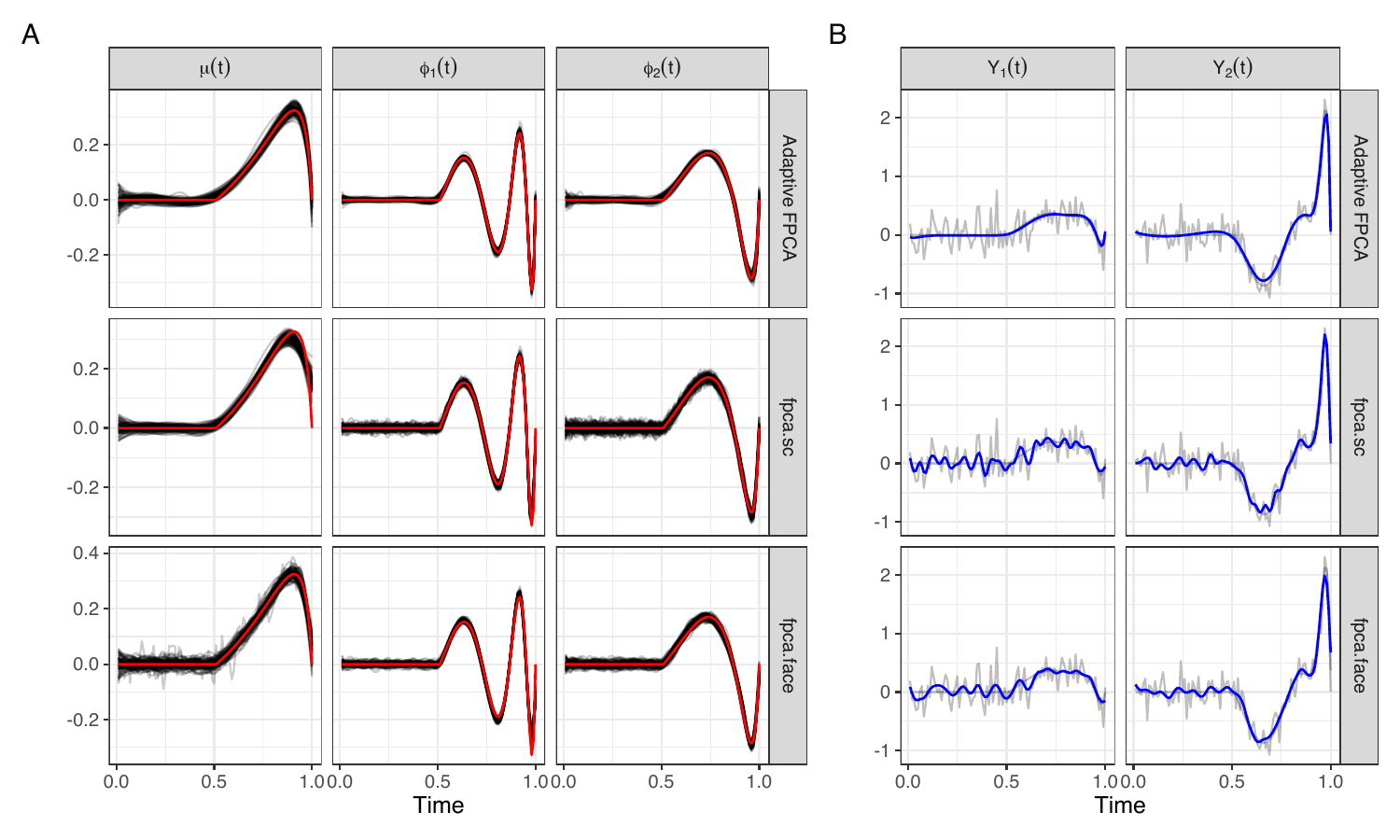} 
     \end{tabular}
     \caption{Simulation results are presented for $I = 25$ and $\sigma^2 = 0.2$. Panel A displays the true data generating curves (in red) overlaid on the estimated data-generated curves from 100 simulations. Panel B showcases two representative curves and their corresponding reconstructions (in blue) obtained by each method. Our proposed method demonstrates better smoothing compared to \texttt{fpca.face()} and \texttt{fpca.sc()}, which exhibit local undersmoothing.}
    \label{fig:curves_estimates}
\end{figure}

Figure \ref{fig:curves_estimates} illustrates the results for simulations with $I = 25$ and $\sigma^2 = 0.2$. Panel A displays the estimates of the true data-generating curves, with columns showing the estimates across the mean $\mu(t)$ and FPCs $\phi_1(t)$ and $\phi_2(t)$. In each panel, black lines shown are estimates from a single simulated dataset; red lines show the true functions. Our method appropriately smooths the constant portion of each function in the first half of the observation window while retaining sharp changes in the second half. \texttt{fpca.sc()} captures the sharp changes but undersmooths over most of the functional domain, resulting in noisy estimates. \texttt{fpca.face()} is similar, but it does not oversmooth the peak in the mean. Panel B shows two randomly selected simulated curves from a single dataset and their reconstructions using all retained FPCs for each method. As expected, our adaptive approach produces reconstructions with appropriate smoothness across the entire time window. \texttt{fpca.face()} and \texttt{fpca.sc()} yield noisier reconstructions, as they struggle to balance complexity and smoothness.

\begin{figure}[htb]
  \centering
     \begin{tabular}{cc}
       \includegraphics[width= 0.8\textwidth]{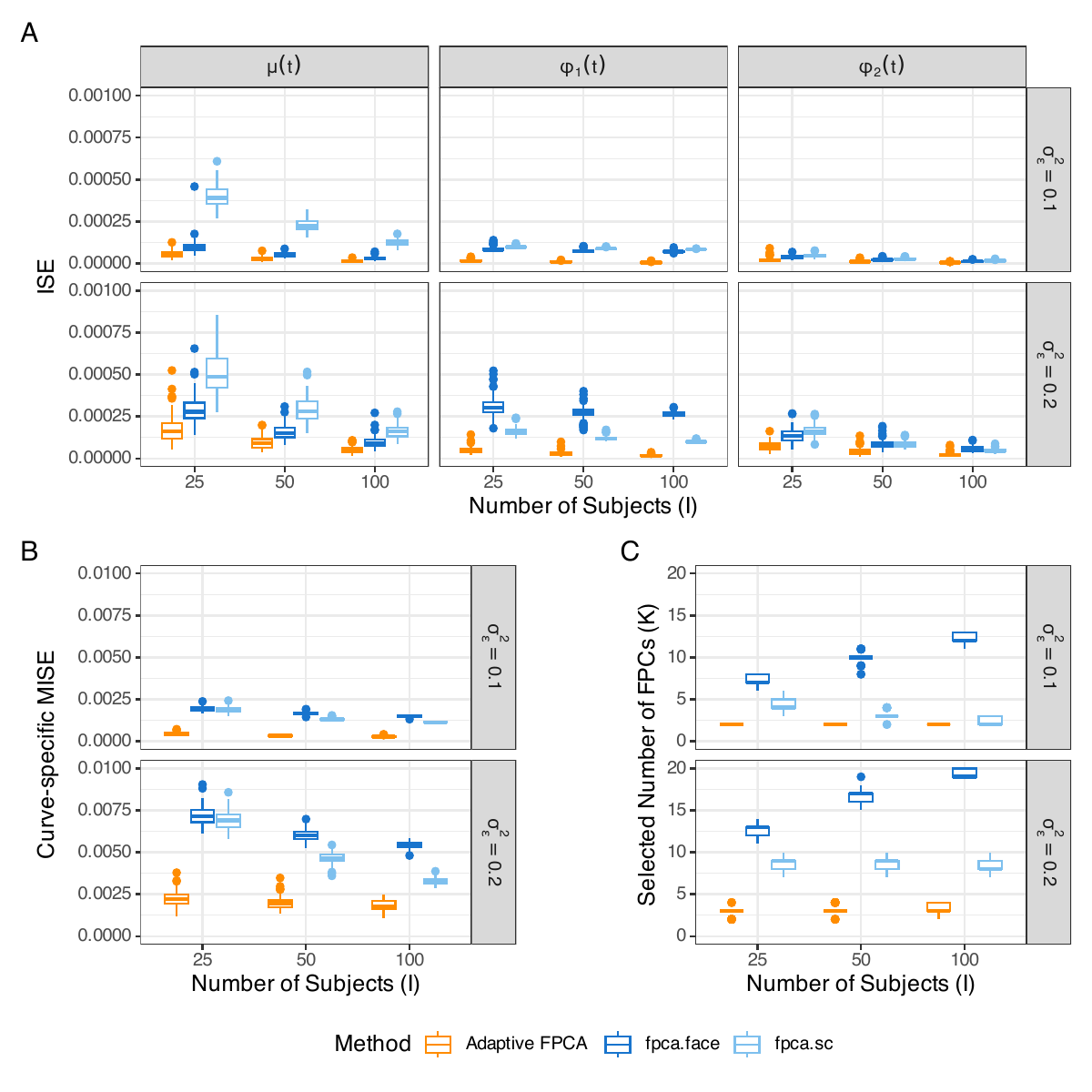} 
     \end{tabular}
     \caption{Panel A shows the accuracy of our proposed method (in orange), and the two standard approaches (in blue) when estimating the true data-generating functions. Panel B is a plot of MISE curve-specific reconstruction. Panel C shows Boxplots of selected FPCS ($K$) using a 99\% PVE threshold}
    \label{fig:MISE_boxplot}
\end{figure}

In Figure \ref{fig:MISE_boxplot}, Panels A-B summarise estimation accuracy across each combination of sample size and noise level. Panel A displays the ISE for the mean $\mu(t)$, and fpcs $\phi_1(t)$, and $\phi_2(t)$. As expected, all methods perform better as the sample size increases and worse as the residual variance increases. These changes are more profound for both \texttt{fpca.sc()} and \texttt{fpca.face()}, as our method provides good accuracy for a range of $I$ when fixing $\sigma^2$. \texttt{fpca.sc()} is less accurate at estimating the mean, as it tends to oversmooth the data compared to \texttt{fpca.face()} and our implementation. \texttt{fpca.face()} and \texttt{fpca.sc()} perform worse than our method when estimating FPCs due to the high wiggliness of both of these components.
These results are consistent with the general pattern observed in Panel A of Figure \ref{fig:curves_estimates}. We present the observation-specific MISE in Panel B. Again, all methods perform better as $I$ increases and worse as the $\sigma^2$ increases. Consistent with the general pattern of panel B Figure \ref{fig:curves_estimates}, our method outperforms both \texttt{fpca.face()} and \texttt{fpca.sc()} across all simulation scenarios.

Panel C of Figure \ref{fig:MISE_boxplot} shows the number of FPCs selected across each simulation setting. Our method retains the correct number of components in the low noise scenarios and usually only one more in the high noise setting. On the other hand, \texttt{fpca.sc()} and \texttt{fpca.face()} generally require more (often many more) components to explain 99\% of the variation; these approaches undersmooth the overall covariance to capture rapidly changing features, and as a result, each component explains less overall variability. Since the number of retained components will affect reconstruction accuracy, we considered a PVE threshold of 95\% and 99.9\% as part of a sensitivity analysis; however, this did not meaningfully change the results observed in this section.

\section{Application Results}
\label{sec:application}

Our scientific goal is derive activation patterns that capture the motor cortex's neural behavior during skilled movements \citep{sauerbrei2020cortical}. We examine a dataset containing firing rates in the motor cortex of a trained mouse reaching for a food pellet in response to an auditory cue. We analyze trial-averaged activation data for 25 neurons over a 1.75-second observation window beginning 0.25 seconds before the cue (Figure \ref{fig:figure1}). We apply adaptive FPCA and compare the results to those obtained using the non-adaptive approach implemented in $\texttt{refund::fpca.sc()}$. We fix the number of splines basis functions to be 40 and retain enough FPCs that explain 99\% of the variability in the data, resulting in 4 and 5 retained components for the adaptive and non-adaptive methods, respectively. 

\begin{figure}[htb]
  \centering
     \begin{tabular}{cc}
       \includegraphics[width= 1\textwidth]{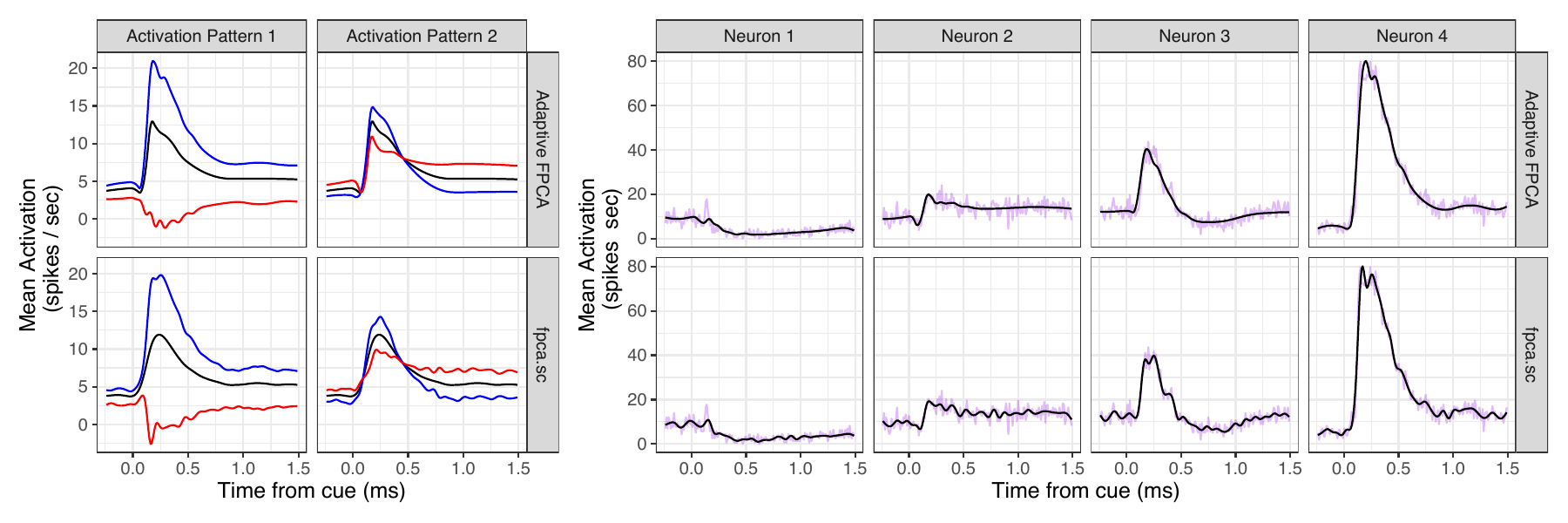} 
     \end{tabular}
     \caption{Application of proposed methodology and $\texttt{refund::fpca.sc()}$ to motivating dataset. Top row displays estimates from Adaptive FPCA. Bottom row shows the same estimates derived using $\texttt{fpca.sc}$. The first two columns shows the variability explained by the first two activation patterns (these plots show the mean activation (in black) plus (in blue) activation pattern times $75^{th}$ score quantile and plus (in red) activation pattern times $25^{th}$ score quantile. The last three columns are the curve reconstructions for Neuron 2-4 in Figure 1.}
    \label{fig:figure4}
\end{figure}

Figure \ref{fig:figure4} displays the first two derived FPCs and curve reconstructions for the four example neurons shown in Figure \ref{fig:figure1}. The top row shows the results from adaptive FPCA. The estimated FPCs are smooth before the cue and in later parts of the reach but show a sharp change at the time of the reach. The first FPC explains 89.0\% of the variance and predominantly captures the amplitude of the initial activation peak in response to the auditory cue. The second FPC explains 6.7\% of the variance and illustrates a contrast comparing activation in $[0, 0.5]$ to activation in $[0.5, 1.5]$: neurons with positive scores on this FPC have above average initial activation and below average later activation in response to the cue, and conversely.

In the right panel, curve reconstructions from our approach demonstrate overall smoothness, yet sharp changes in activation are modeled appropriately. Neurons 1 and 2 are largely flat, although neuron 1 is somewhat inhibited after the auditory cue and neuron two shows elevated activity. Neurons 3 and 4 increase sharply; neuron three then momentarily dips whereas neuron 4 gradually returns to baseline. In contrast to our method, and as displayed on the bottom row, non-adaptive FPCA captures the sharp increase effectively but undersmooths activity before the cue and in later times of the reach. Our findings are consistent with our simulation analyses in Section \ref{sec:sim}. 

\section{Discussion}
\label{sec:disc}

The adaptive smoothing functional principal component analysis method proposed in this manuscript is an innovative tool in functional data analysis and is the appropriate analysis for our motivating dataset. We first developed a fast adaptive scatterplot smoothing technique and then incorporated that approach into a likelihood-based FPCA framework. The adaptive smoothing technique draws on an explicit connection to adaptive ridge penalties, establishing a helpful analog to previous work in non-adaptive scatterplot smoothing and motivating a data-driven way to estimate tuning parameters. The proposed adaptive scatterplot smoothing and adaptive functional principal component analysis methods are both relevant to contexts beyond our specific application. We have developed publicly-available software to facilitate the use of these tools. 

In the analysis of our motivating dataset, we estimated activation patterns that have the locally-varying degrees of smoothness required to model sharp changes in motor cortex activation in response to a cue that triggers a reaching motion and the smoothly-varying changes observed elsewhere in the observation window. The patterns have interpretable explanations in the context of the reaching experiment, and yield good fits to the activation of the individual neurons. Our results correctly model the resting state of the motor cortex and the sharp changes that occur in activation patterns after a mouse reacts to an auditory cue and reaches for a pellet. Because activation patterns derived through dimension reduction methods are commonly used in analyses that connect later paw position to brain activity, our work may lead to better data processing and more robust overall analysis pipelines. 

Several directions for future work remain. Our approach adaptively smooths each FPC separately. In the context of our motivating data, sharing information about the local degree of smoothness or including prior information about the timing of the auditory cue would provide a mechanism to incorporate important information across components. Alternatively, developing an adaptive bivariate smoothing approach for covariance surfaces could also lead to FPC estimates with co-located areas of varying degrees of smoothness. We have developed methods for Gaussian data and modeled the trial-averaged neural activation. Extending the adaptive FPCA method to generalized or multilevel settings could provide a way to model the data without the need to average across trials and to handle count or binary activation. Lastly, our focus was on estimating activation patterns using FPCA; in the future, including our fast adaptive smoothing approach in functional regression methods may improve the ability to estimate coefficient functions that exhibit locally-varying degrees of smoothness.


\section{Acknowledgments}
\label{sec:ack}

This work was supported by Award R01NS097423-01 from the National Institute of Neurological Disorders and Stroke. 


\setstretch{1.2}

\bibliographystyle{jasa} 
\bibliography{biblio}

\setstretch{1.8}


\end{document}